\documentclass[nofootinbib, floatfix, notitlepage,twocolumn]{revtex4-1}

\usepackage{color}
\usepackage{amsmath,amsfonts,amssymb}
\usepackage{physics}
\usepackage{bm}
\usepackage{mathrsfs}
\usepackage{graphicx}
\usepackage{centernot}
\usepackage[english]{babel} 
\usepackage{hyperref} 
\usepackage[normalem]{ulem}
\usepackage{textcomp}

\usepackage{bbold}
\usepackage[utf8]{inputenc}

\usepackage{float}

\begin{document}

\title{Gravitational Wave Energy Emitted in the Head-On Collision of Two Black Holes}

\author{Nesibe Derin Sivrioglu and Robert R.\ Caldwell} 
\affiliation{Department of Physics and Astronomy,
Dartmouth College, Hanover, NH 03755, USA}

\begin{abstract}
What is the spectrum of gravitational radiation produced by the head-on collision of two equal-mass black holes? The emission is dominated by low frequency bremsstrahlung, producing a flat energy spectrum. But where does the spectrum turn over? We propose that the lowest quasinormal mode of the final black hole marks the end of the low-frequency domain. The result is an analytic model of the total emitted energy as a function of the black hole velocity in the center of mass frame. With no free parameters, the model predicts that in the speed-of-light limit, $13.8\%$ of the total initial energy is emitted in gravitational radiation, in good agreement with numerical relativity. This result also enables calculation of the nonlinear contribution to the memory, a persistent distortion of the spacetime after passage of the gravitational wave burst. Advances in numerical relativity simulations will enable tests of our model for increasingly relativistic speeds, providing insight into this extreme collision. 

\end{abstract}

\date{\today} 
\maketitle

The head-on collision of two black holes presents a challenging problem in gravitational physics. Traveling at relativistic speeds, the black holes are Lorentz-contracted into thin disks that ultimately merge in what must be one of the most violent processes imaginable. The collision is a testbed for theory as well as numerical relativity. What is the spectrum of gravitational radiation from this collision, what is the mass of the final black hole, and how do these depend on the speed of the two initial black holes? These are the questions we aim to address.


In seeking the total energy in gravitational radiation, several approaches that circumvent the nonlinear nature of such an interaction have been explored. Hawking famously applied the area-increase theorem to any process by which two black holes combine to form a third. The result is an upper bound of $29\%$ for the energy in gravitational radiation \cite{Hawking:1971tu}. Penrose, considering infinitely-boosted Schwarzschild black holes, found an apparent horizon at the moment of collision that allows one to put the same $29\%$ bound for the energy in gravitational radiation \cite{penrose}. Smarr applied a low frequency approximation, the ``zero frequency limit" (ZFL), to a classical, linearized calculation of the gravitational wave spectrum generated by scattering point particles \cite{Smarr:1977fy}. This matched the earlier result by Weinberg \cite{Weinberg:1965nx}, who considered the soft gravitons attached to external legs of any scattering diagram. Both methods predicted a constant differential energy spectrum $dE/d\omega$, requiring an upper frequency cutoff, $\omega_c$, to determine the total energy in gravitational radiation. In the ultrarelativistic regime, D'Eath and Payne \cite{DEath:1992mef,DEath:1992nmz,DEath:1992plq} treated the head-on, speed-of-light collision of black holes perturbatively, giving an estimate of $16\%$ for the fraction of the total energy in gravitational radiation. Most recently, progress in numerical relativity has enabled high resolution simulations of head-on collisions \cite{Sperhake:2008ga, Healy:2024lhl,PhysRevD.98.084053}, obtaining the full gravitational waveform and energy spectrum. In particular, the results of numerical simulations are used to calibrate the cutoff, $\omega_c$, that determines the overall scale of the ZFL prediction for the energy in gravitational radiation. By extrapolating these results for finite relativistic $\gamma$ factor to $\gamma \to \infty$ using the ZFL model, an efficiency of approximately $14\%$ is obtained \cite{Sperhake:2008ga,PhysRevD.94.104020}.


In this article, we propose that the cutoff is merely the lowest quasinormal mode of the final, stationary black hole. The cutoff is determined with no free parameters, and results in a better fit to simulation data than the standard ZFL model. The new model predicts that $13.8\%$ of the total initial energy is emitted in gravitational radiation, in excellent agreement with numerical relativity.


To begin, we consider the interaction between a set of unbound particles with uniform, initial (final) momenta $p'$ ($p''$). 
We examine the process on time and length scales that are large compared to the details of the scattering interaction.
It is then shown that at late times, the change in the gravitational field,
\begin{equation}
    \Delta h^{TT}_{ij} = \frac{4}{R} \left( \sum_{out}\frac{p''^i p''^j}{p''\cdot \hat q} -\sum_{in}\frac{p'^i p'^j}{p'\cdot \hat q} \right)^{TT},
    \label{eqn:hij}
\end{equation}
represents the waveform of low frequency gravitational waves. We use units where Newton's constant and the speed of light are set to unity, $R$ is the distance from source to observer, $\hat q = (1,\hat n)$ is a null 4-vector pointing in the direction $\hat n$, and superscript $TT$ indicates the transverse-traceless component is taken. It is straightforward to write the gravitational wave angular power as
\begin{equation}
    \frac{dL}{d\hat n} = \frac{R^2}{16 \pi} \sum_P \dot h_P^2
\end{equation}
where $P$ is the polarization and $h_P = \frac{1}{2} h_{ij}^{TT} \epsilon_P^{ij}$. 
Integrating the luminosity over all time, we can then use Parseval's theorem to express the energy in the frequency domain. Taking the low frequency limit we arrive at an expression for the gravitational wave energy spectrum
\begin{equation}
    \frac{d^2 E}{d\omega\, d\hat n} \bigg|_{ZFL}\equiv \frac{R^2}{16 \pi^2} 
    \sum_P \left( \Delta h_{P}\right)^2, \label{eqn:B}
\end{equation}
where $\Delta h_{P}$ is the late time change in the gravitational field given by Eq.~\eqref{eqn:hij}. In this set up the energy spectrum is 
independent of frequency.

We may now specialize to the case of two colliding black holes. The initial momenta are $p' = \gamma m(1, \pm v \hat z)$. After the collision there are no outgoing particles, in the center of mass  frame, and so $p''=0$. Using Eq.~(\ref{eqn:B}), we find the head-on collision produces a differential spectrum
\begin{equation}
    \frac{d^2E}{d\omega \, d\hat n} = \frac{\gamma^2 m^2 v^4}{\pi^2 }\frac{\sin^4\theta}{(1 - v^2 \cos^2\theta)^2}. \label{eqn:smarr}
\end{equation}
Integrating over all directions, we find
$dE / d\omega = M^2 f(\gamma)/\pi$ where
\begin{equation}
f(\gamma) =  \frac{1+2 \gamma^2}{2 \gamma^2}+ \frac{(1-4\gamma^2)\ln(\gamma + \sqrt{\gamma^2-1})}{2 \gamma^3 \sqrt{\gamma^2-1}} \label{eqn:Ef}
\end{equation}
and $M \equiv 2 \gamma m$ is the total energy of the system. The function $f$ interpolates between $0$ and $1$ as $\gamma$ goes from $1$ to $\infty$. To determine the total energy in gravitational radiation, we integrate the differential spectrum up to some cutoff, $\omega_c$.


In previous literature, the product $\omega_c M$ has been treated as a constant, free parameter to be determined by numerical experimentation \cite{Berti:2010ce}. However, the cutoff represents the limit of the domain of validity of the low-frequency approximation, which is set by the time or length scales of the ``scattering" interaction. 


Here, we propose that the fundamental scale must be the real part of the lowest quasinormal mode of the final black hole, corresponding to the ringdown frequency. At all higher frequencies, the physics is determined by a superposition of quasinormal modes and the energy is exponentially suppressed. The  energy in  gravitational radiation is
\begin{equation}
    \frac{E}{M} = \frac{\omega_c M}{\pi} f(\gamma).
\end{equation}
The dominant quasinormal mode of the final black hole can be found as
\begin{equation}
    \omega_{QNM} =\frac{\Omega}{M_{f}},
    \label{eqn:qnms}
\end{equation}
where $\Omega=0.3737$ is the real part of the $l=2$ mode of a Schwarzschild black hole in units of the black hole mass (e.g. Refs.~\cite{Chandrasekhar:1975zza,Kokkotas:1999bd}, Table 1), and $M_{f}=M-E$ is the mass of the black hole resulting from the head-on collision. Equating the cutoff frequency to the lowest quasinormal mode frequency, we achieve 
\begin{equation}
    \omega_{c} M = {\Omega}/{(1-\frac{E}{M})} \rightarrow {\Omega}/{(1-\frac{\omega_{c} M}{\pi}f(\gamma))},
\end{equation}
which may be solved for $\omega_c$, 
\begin{equation}
    \omega_{c} M =  \frac{\pi}{2f(\gamma)} \left( 1 - \sqrt{1 - \frac{4}{\pi} \Omega f(\gamma)} \right),
    \label{eqn:cutoff}
\end{equation}
whereby the energy is
\begin{equation}
    \frac{E}{M}=\frac{1-\sqrt{1-\frac{4}{\pi}\Omega f(\gamma)}}{2}.
    \label{eqn:energyresult}
\end{equation}
This is our main result, consisting of a new model of the dependence of energy on the relativistic $\gamma$ factor, which yields $E/M = 0.138$ in the limit $\gamma \to \infty$. The conventional ZFL calibrated to the same value at $\gamma \to \infty$ behaves as $E/M =\frac{1}{2}(1 - \sqrt{1-4\Omega/\pi}) f(\gamma)$, clearly different from Eq.~(\ref{eqn:energyresult}).

\begin{figure}[t]
    \centering
    \includegraphics[width=0.95\linewidth]{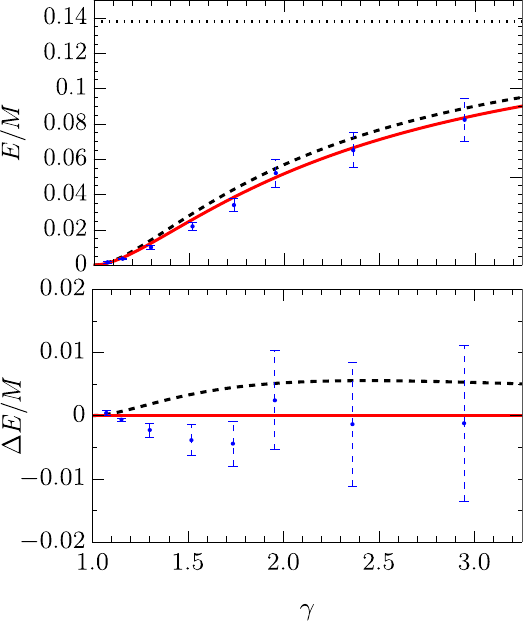}
    \caption{
    (top) Energy radiated in gravitational waves in the head-on collision of two equal-mass black holes, as a fraction of the total energy, $E/M$ versus the relativistic $\gamma$ factor. Solid, red is our proposed model, black dashed is the standard ZFL, and blue dots with error bars are simulation data \cite{Sperhake:2008ga}. As $\gamma \to \infty$ the proposed model asymptotes to $E/M \to 0.138$, shown by the dotted black line. (bottom) Same as top, relative to new model. The new model provides a significantly better fit to the data than the standard ZFL. 
    }
    \label{fig:energyresult}
\end{figure}

Our new model, Eq.~(\ref{eqn:energyresult}) is shown in Fig.~\ref{fig:energyresult}. For comparison, we also show the standard ZFL model and recent numerical results \cite{Sperhake:2008ga}. The new model provides a substantially better fit to the simulation data, yielding $\chi^2=17$ for eight data points without any free parameters. By contrast, the standard ZFL treatment yields $\chi^2=62$ despite containing one free parameter.
%

The figure also helps us understand recent findings that the ZFL appears to yield a poor fit to the data \cite{Berti_2025}: a fit to the low $\gamma$ data points yields a different asymptotic value at $\gamma\to\infty$ than a fit to the higher $\gamma$ data points. The shape of the standard ZFL simply does not match the simulation data, as seen in the figure. As numerical relativity simulations continue to improve, extending to larger values of $\gamma$ and yielding smaller uncertainties, we expect increasingly stringent tests of our model, making it distinguishable from other analytical descriptions.

\begin{figure}[H]
    \centering
    \includegraphics[width=0.95\linewidth]{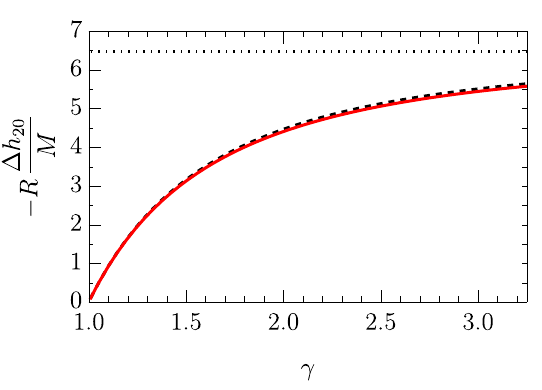}
    \caption{
    The $l=2$ mode of the gravitational wave memory. Dashed black is the linear memory, solid red includes both linear and nonlinear memory.     The asymptotic, $\gamma \to \infty$ value is shown by the dotted black line.}
    \label{fig:h2all}
\end{figure}

We can use our new model to answer further questions regarding the late-time change in the gravitational field in the form of memory. Recall that an additional contribution to the gravitational memory arises due to the nonlinearity of Einstein's equations, as shown by Christodoulou \cite{PhysRevLett.67.1486}. Thorne later interpreted this ``nonlinear memory" as arising from the contribution of gravitational waves \cite{PhysRevD.45.520}. To describe this contribution, we extend our treatment given in Eq.~\eqref{eqn:hij} to include emitted gravitational waves carrying four-momentum:  $p''=\frac{d E}{d \hat{n}''}(1,\hat{n}'')$. Here, we use the results of our new model: integrating the differential spectrum in Eq.~\eqref{eqn:smarr} up to a cutoff frequency $\omega_{c}(\gamma)$ given by Eq.~\eqref{eqn:cutoff}. The determination of $\omega_c(\gamma)$ then allows us to describe the angular distribution of energy for low frequencies. When applied to Eq.~\eqref{eqn:hij}, the total contribution to the late time change in the gravitational field becomes:
\begin{equation}
    \Delta h_+ = \frac{4}{R} \left[ -\frac{\gamma m v^2\sin^2\theta}{1 - v^2 \cos^2\theta} + \int d\hat n'' \, \frac{dE}{d\hat n''} \frac{\hat n_i'' \hat n_j'' }{1 - \hat n \cdot \hat n''} \frac{1}{2}\epsilon_+^{ij} \right]
\end{equation}
where $\Delta h_\times=0$ by symmetry. The first term arises from the mechanical motion of the gravitating sources and corresponds to the linear memory. The second term arises from the gravitational radiation itself and corresponds to the nonlinear memory. After a brief calculation, the memory projected onto the $\ell=2,\, m=0$ spin-2 spherical harmonic is given by
\begin{eqnarray}
    \Delta h_{+L} \frac{R}{M}|_{20} &=& 
    \sqrt{\frac{10 \pi}{3}}a(v) \\
    \Delta h_{+NL} \frac{R}{M}|_{20} &=& 
    \sqrt{\frac{5 \pi}{24}}\frac{1 - \sqrt{1 - \frac{4}{\pi}\Omega f(\gamma)}}{f(\gamma)} b(v).
\end{eqnarray}
where $a(v) = (3 v - 5 v^3 -3(1-v^2)^2\tanh^{-1} v)/v^3$ and $b(v) = (1-v^2)(v(v^2-15) + (15 - 6 v^2 - v^4)\tanh^{-1}v)/v^3$. The overall amplitude of the nonlinear contribution is surprisingly small, roughly $1\%$ of the linear memory at peak, and vanishes in both the non-relativistic and ultra-relativistic limits. The latter occurs because the angular distribution of energy becomes isotropic in the limit $\gamma \to \infty$  \cite{Smarr:1977fy}, and thereby does not contribute to the nonlinear strain. The memory for the $l=2$ mode is shown in Fig.~\ref{fig:h2all}. These waveforms manifest as a DC offset of the gravitational field that persists after the burst of high frequency gravitational waves has passed by \cite{Caldwell:2025tfu}.

We note that the energy, memory, and cutoff are connected via Eq.~(\ref{eqn:B}) for interactions dominated by low frequencies. This opens the possibility to use late-time data such as the final black hole mass and the memory displacement to approximate the fundamental quasinormal mode frequency, a complementary method to a spectroscopic analysis \cite{Berti:2005ys}.

In this brief note we have presented a new model of the dominant, low-frequency gravitational radiation emitted in the head-on-collision of two equal mass black holes. The model consists of self-consistently equating the high frequency cutoff with the leading quasinormal mode of the final, stationary black hole. The energy of radiation at higher frequencies is exponentially suppressed. Thus, we find that Eq.~(\ref{eqn:energyresult}) gives the approximate energy emitted in gravitational waves as a fraction of the total energy; $13.8\%$ in the limiting case of two ultrarelativistic black holes. We await the results of forthcoming numerical relativity simulations \cite{Berti_2025} to put our new model, illustrated in Figs.~\ref{fig:energyresult}-\ref{fig:h2all}, to the test. Although such a head-on collision is unlikely to occur in nature, extreme cases often provide insight that translates to more probable situations. For future work, natural extensions to consider are the energy spectrum and cutoff for unequal mass \cite{Berti:2010ce}, off-axis collisions \cite{Healy:2024lhl,Shibata:2008rq}, and fly-by encounters \cite{Dandapat:2023zzn}.

\acknowledgments
We thank Emanuele Berti for useful conversations.

\vfill
 
\bibliography{main}

\vfill

\end{document}